\documentclass{aa}

\usepackage{graphicx}
\usepackage{txfonts}

\usepackage{natbib}
\usepackage{times}

\newcommand {\asp}{\mbox{$.\!\!^{\prime\prime}$}}

\newcommand{\etal}{et al.,~}
\newcommand{\msun}{{\,\rm M}_{\odot}}

\newcommand{\kms}{\,{\rm km\,s}^{-1}}

\newcommand{\nht}{\ifmmode {{\rm NH}_3} \else {NH{\bas 3}} \fi}
\newcommand{\tco}{\ifmmode {^{13}{\rm CO}} \else {$^{13}{\rm CO}$}\fi}
\newcommand{\dco}{\ifmmode {^{12}{\rm CO}} \else {$^{12}{\rm CO}$}\fi}
\newcommand{\cdo}{\ifmmode {{\rm C}^{18}{\rm O}} \else {${\rm C}^{18}{\rm O}$}\fi}

\begin{document}

\title{\bf Rotating molecular outflows: the young T Tauri star in CB\,26
      \thanks{Based on observations carried out with the IRAM Plateau
      de Bure Interferometer.  IRAM is supported by INSU/CNRS
      (France), MPG (Germany) and IGN (Spain). Also based on
      observations collected at the Centro Astron\'{o}mico Hispano
      Alem\'{a}n (CAHA) at Calar Alto, operated jointly by the
      Max-Planck Institut f\"ur Astronomie and the Instituto de
      Astrof\'isica de Andaluc\'ia (CSIC).}}

\author{Ralf Launhardt \inst{1},
Yaroslav Pavlyuchenkov \inst{1,2},
Fr\'ed\'eric Gueth \inst{3},
Xuepeng Chen \inst{1},
Anne Dutrey \inst{4,5},
St\'ephane Guilloteau \inst{4,5},
Thomas Henning \inst{1},
Vincent Pi\'etu \inst{3},
Katharina Schreyer \inst{6},
Dmitry Semenov \inst{1}}
\offprints{Ralf Launhardt, \email{rl@mpia.de}}
\institute{Max-Planck-Institut f\"ur Astronomie, K\"onigstuhl 17, D-69117
Heidelberg, Germany 
\and
Institute of Astronomy, Russian Academy of Sciences, Pyatnitskaya 48, Moscow, 109117, Russia
\and
IRAM, 300 rue de la piscine, F-38406 Saint Martin d'H\`eres, France
\and
Universit\'e Bordeaux 1, Laboratoire d'Astrophysique de Bordeaux (LAB), France
\and
CNRS/INSU - UMR5804; BP 89, F-33270 Floirac, France
\and
Astrophysikalisches Institut und Universit\"ats-Sternwarte,
Schillerg\"asschen 2-3, D - 07745 Jena, Germany}
\date{Received 20 August 2008 / Accepted 18 November 2008}



  \abstract
   {The disk-outflow connection is thought to play a key role in extracting excess angular
    momentum from a forming proto-star. Though jet rotation has been observed in a few
    objects, no rotation of molecular outflows has been unambiguously reported so far.}
   {We report new millimeter-interferometric observations of the 
    edge-on T\,Tauri\,star -- disk system in the isolated Bok globule CB\,26.
    The aim of these observations was to study the disk-outflow relation in this 
    1\,Myr old low-mass young stellar object.}
   {The IRAM PdBI array was used to observe  
   $^{12}$CO(2--1) at 1.3\,mm in two configurations, resulting in
   spectral line maps with 1.5\arcsec\ resolution. 
   We use an empirical parameterized steady-state outflow model combined with 
   2-D line radiative transfer calculations and $\chi^2$-minimization in parameter space 
   to derive a best-fit model and constrain parameters of the outflow.}
   {The data reveal a previously undiscovered collimated bipolar molecular 
    outflow of total length $\approx 2000$\,AU, escaping perpendicular to the plane 
    of the disk. We find peculiar kinematic signatures that suggest the outflow 
    is rotating with the same orientation as the disk.
    However, we could not ultimately exclude jet precession or two misaligned flows 
    as possible origin of the observed peculiar velocity field. 
    There is indirect indication that the embedded driving source is a
    binary system, which, together with the youth of the source, could
    provide the clue to the observed kinematic features of the
    outflow.}
    {CB\,26 is so far the most promising source to study the rotation of a 
    molecular outflow. Assuming that the outflow is rotating, we compute and
    compare masses, mass flux, angular momenta, and angular momentum
    flux of disk and outflow and derive disk dispersal timescales of
    $0.5\ldots1$\,Myr, comparable to the age of the system.}
%
%

\authorrunning{Launhardt \etal}

\titlerunning{Rotating molecular outflow in CB\,26}

\maketitle

\section{Introduction} \label{sec_intro}

A comprehensive understanding of the complex relations between a
collapsing molecular cloud core, the forming young (proto-)star, its
circumstellar disk, and the expelled bipolar jets is a necessary
prerequisite to understand the formation of stars and planetary
systems. In particular, the initial amount, evolution, and
re-distribution of angular momentum appear to be key factors in many
of the processes which determine the protostellar fragmentation, the
final stellar mass(es), and the existence and morphology of planetary
systems.

Rotation of protostellar jets and molecular outflows about their
symmetry axis is believed to be a key mechanism to remove angular
momentum from forming protostars and, thus, at least partially, to
solve the angular momentum problem in star formation
\citep[e.g.,][]{2007prpl.conf..231R}.  As a consequence, many groups
have searched for rotation signatures in jets and outflows.  
Indeed, during the past few years, 
radial velocity\footnote{Throughout this paper, 
we use the term "radial velocity" when we refer to the 
line-of-sight velocity in the coordinate system of the observer. 
For the radial component of the gas motion in the coordinate system 
of the source (i.e., gas moving radially away from the source), we use the term 
"radial outflow velocity".} 
gradients have been observed across a number of protostellar jets 
using high-resolution HST spectra
\citep[]{bachiotti2003,coffey2004,woitas_etal2005,coffey2007}.  These
were interpreted as jet rotation tracing the helicoidal structure of
the magnetic field in accretion-ejection structures.  As these
observations are very difficult and the rotation signatures are weak,
other groups have also searched for rotation signatures in molecular
outflows.  While \citet{Pety_etal2006} did not detect outflow rotation
in HH\,30, \citet{Lee_etal2007} report radial velocity gradients across SiO
knots in HH\,211 which they interpret as jet rotation. However, recent
PdBI observations could not confirm this claim (Gueth et al. in prep.).

In order to study the chemical composition, kinematics, and evolution
of protoplanetary disks, we started the CID project ("Chemistry In
Disks"), a molecular line survey with the IRAM Plateau de Bure
Interferometer (PdBI) of well-known disks such as those surrounding
LkCa\,15, MWC\,480, or DM\,Tau \citep{dutrey_etal2006}.  Our main goal
is to map, at high angular resolution, various molecular lines with
enough sensitivity to derive molecular abundance variations versus
radius.  For the youngest source in our sample, CB\,26, we also mapped
the molecular outflow with the goal to study the disk\,--\,outflow
relation.

CB\,26 (L\,1439) is a small cometary-shaped Bok globule located
$\approx$\,10$^\circ$ north of the Taurus-Auriga dark cloud, at a
distance of $\approx$\,140\,pc.  A dense core with signatures of star
formation is located at the south-west rim of the globule
\citep{lau_hen1997}.  OVRO observations of the mm dust continuum
emission and of the $^{13}$CO\,(1-0) line have revealed a nearly
edge-on circumstellar disk of radius 200\,AU with Keplerian rotation
\citep{lau_sar2001}, surrounding a very young (obscured) low-mass T
Tauri star.  It is associated with a small bipolar near-infrared (NIR)
nebula bisected by a dark extinction lane at the position and
orientation of the edge-on disk \citep{stecklum_etal2004}.  The source
is surrounded by an optically thin asymmetric envelope with a well-ordered
magnetic field directed along P.A.\,$\sim$\,25\degr\
\citep{henning_etal2001}.  Furthermore, an Herbig-Haro object
(HH\,494) was identified by H$\alpha$ and S[II] narrow-band imaging,
6.15 arcmin northwest of CB\,26 at P.A.\,=\,$145$\degr\
\citep{stecklum_etal2004}.  The HH object is thus perfectly aligned
with the symmetry axis of the disk and the bipolar nebula.

Here, we present new interferometric molecular line data of CB\,26 and
report the discovery of a small, well-collimated bipolar molecular outflow.
The outflow has a peculiar velocity structure that suggests it is 
rotating with the same orientation as the disk.

\section{Observations and data reduction} \label{sec_obs}

Observations of CB\,26 were carried out with the IRAM PdBI in November
(D configuration with 5 antennas) and December 2005 (C configuration
with 6 antennas). The two array configurations provided baseline
lengths in the range 16\,--\,175\,m.  The phase center was at
$\alpha_{2000} = 04^h59^m50.74^s$, $\delta_{2000} =
52^{\circ}04^{\prime}43.80^{\prime\prime}$.  Two receivers were used
simultaneously and tuned single side-band (SSB) to the
HCO$^{+}$\,(1-0) line at 89.188526\,GHz and the $^{12}$CO\,(2-1) line
at 230.537984\,GHz, respectively, 
adopting the systemic velocity\footnote{All radial velocity 
values in this paper refer to the Local Standard of Rest (LSR).} 
of CB\,26, $v_{\rm LSR}\,=\,5.5$\,km\,s$^{-1}$. 
In this paper, we present only the
$^{12}$CO\,(2-1) data on the molecular outflow. A more detailed
analysis of the disk structure, that includes other molecular line
data, will be presented in a forthcoming paper.

\begin{figure*}
\begin{center}
%
\includegraphics[width=1.0\textwidth]{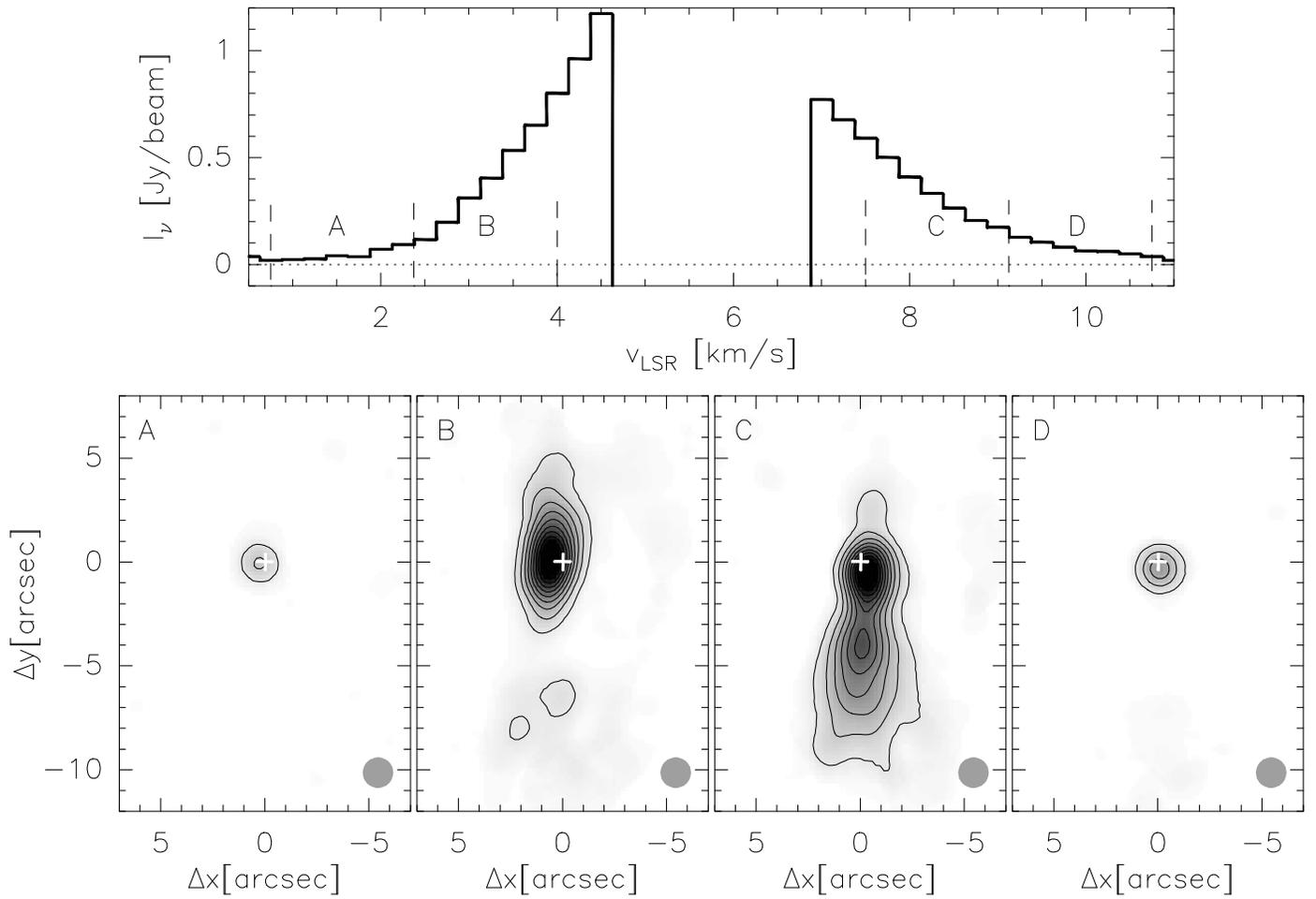}
\caption{\label{fig_chanmap}
 $^{12}$CO\,(2--1) spectrum and binned channel maps of CB\,26. 
Top panel: average $^{12}$CO\,(2--1) spectrum of the central $\pm$2\arcsec.
The central velocity channels between 4.75 and 6.75\,km/s were masked out 
because they are dominated by the extended envelope. The integration ranges for the four 
binned channel maps (bottom panels) are indicated by vertical dashed lines and marked A, B, C, and D.
Bottom panels: binned $^{12}$CO\,(2--1) channel maps of CB\,26, rotated by 30\degr, 
with integration ranges 
indicated by dashed lines in the top panel. Contour levels start at 0.15\,Jy/beam\,km/s. 
The white cross marks the center of the disk. 
The $^{12}$CO synthesized beam size is indicated as grey ellipses in the lower left corners. 
The two outer maps show the high-velocity Keplerian wings 
of the inner disk. The two inner maps are dominated by the blue (upwards) and red (downwards) 
outflow lobes, respectively.}
\end{center}
\end{figure*}

The quasar 0355+508 served as phase calibrator to determine the
time-dependent complex antenna gains.  The correlator bandpass was
calibrated on 3C454.3 and 3C273, and the absolute flux density scale
was derived from observations of MWC\,349.  The flux calibration
uncertainty is estimated to be $\leq$20\%.  The primary beam size at
1.3\,mm was 22\arcsec.  The SSB system temperature at 230\,GHz was in
the range 250--450\,K.  Total bandwidth and effective channel spacing
for the $^{12}$CO\,(2-1) observations where 12\,MHz (16\,km\,s$^{-1}$)
and 32\,kHz (0.25\,km\,s$^{-1}$), respectively.  Zero-spacing data
were obtained in September 2006 with the IRAM 30\,m telescope at Pico
Veleta in Spain.  They were combined with the interferometric data to
help the deconvolution. However, the central velocity channels between
4.75 and 6.75\,km/s, 
which are completely dominated by the extended
envelope, could not be well-restored and were therefore masked out for
the subsequent analysis (Fig.\ \ref{fig_chanmap}).

The data were calibrated and imaged using the
GILDAS\footnote{http://www.iram.fr/IRAMFR/GILDAS} software.  Channel
maps were obtained with natural $uv$-weighting, resulting in an
effective angular resolution of 1.47\arcsec\ (synthesized beam FWHM).
The r.m.s. in the channel maps is 20\,mJy/beam.

\begin{figure*}
\begin{center}
%
\includegraphics[width=1.0\textwidth]{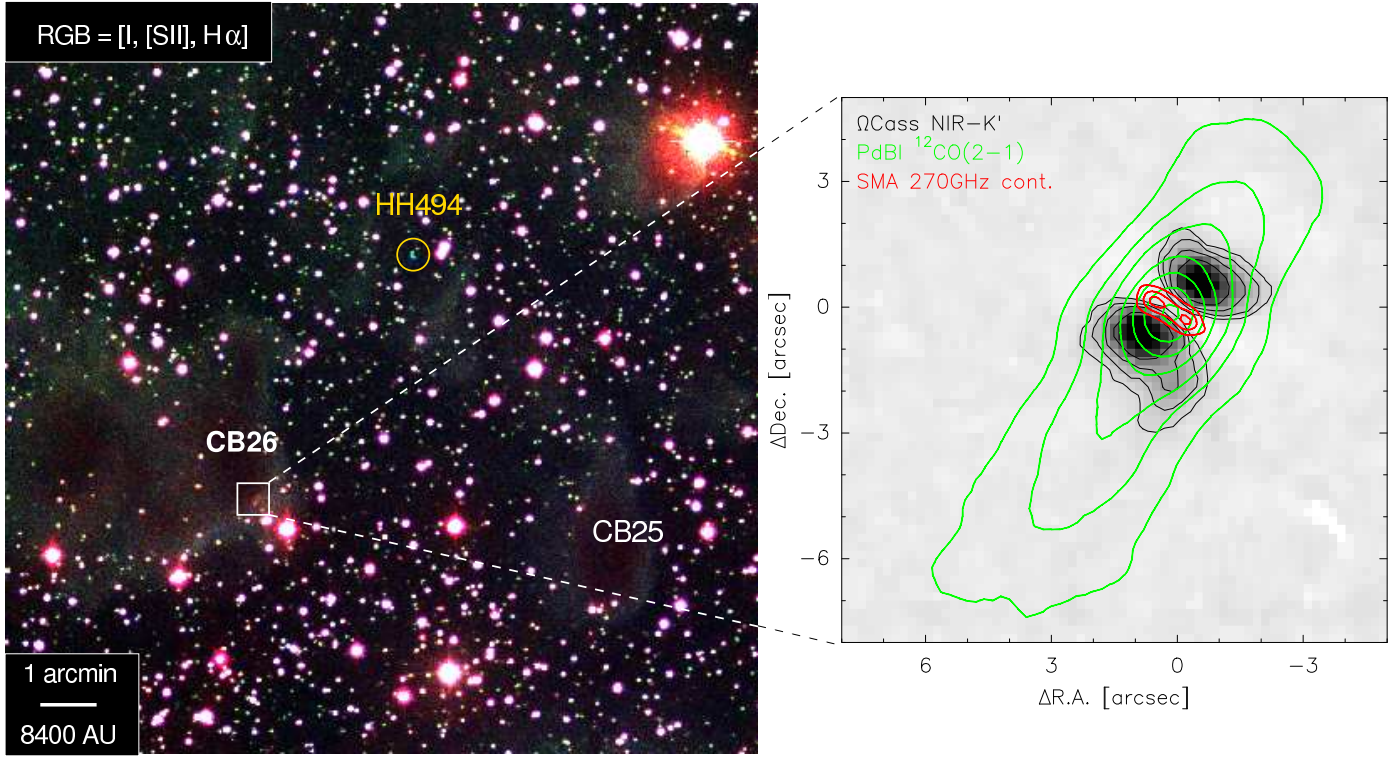}
\caption{\label{fig_overview}
  Overview of the CB\,26 region. 
 The optical true-color image (left) is based on wide-field H$\alpha$ (blue), [SII] (green), 
 and I--band (red) frames \citep{stecklum_etal2004}. The globules CB\,25 and CB\,26 as well as 
 Herbig-Haro object HH\,494 are marked in the image. The zoom panel (right) shows a NIR K-band image 
 of the bipolar reflection nebula (grey-scale, 0\asp6 resolution), 
 overlaid with contours of the SMA 1.1\,mm dust continuum 
 emission from the disk (red, contours at 20, 38, and 55\,mJy/beam) 
 and the integrated $^{12}$CO(2--1) emission (0.6 to 12.5 km/s) from the bipolar molecular 
 outflow (green, contours at 0.5, 1,2, 1.9, 2.9, $\ldots$\ Jy/beam\,km/s). 
 Beam sizes are shown in Figs.\,\ref{fig_chanmap} and \ref{fig_12covelfield}.
 The reference position is $\alpha_{2000} = 04^h59^m50.74^s$, 
                           $\delta_{2000} = 52^{\circ}04^{\prime}43.80^{\prime\prime}$.
}
\end{center}
\end{figure*}

New NIR K$^{\prime}$-band images at 0\asp6 seeing and with 0\asp2
pixel scale were obtained in January 2008 with the Omega Cass camera
\citep{Lenzen_etal1998} at the 3.5\,m telescope of the Centro
Astron\'{o}mico Hispano Alem\'{a}n (CAHA) at Calar Alto in
Spain. These data are used here as complementary information to
illustrate the morphology of the bipolar reflection nebula with
respect to the edge-on circumstellar disk (Fig.\,\ref{fig_overview}).
Additional SMA interferometric data at 270\,GHz to which we refer here
(see also Fig.\,\ref{fig_overview}) will be presented in a forthcoming
paper (Launhardt et al. in prep.).


\section{The $^{12}$CO\,(2--1) maps} \label{sec_COres}

The $^{12}$CO\,(2--1) channel maps and integrated intensity maps of CB\,26
reveal a small, well-collimated, slightly asymmetric bipolar molecular
outflow oriented perpendicular to the plane of the disk at
P.A.\,=\,$147\pm 3$\degr.  
This is illustrated in Fig.\,\ref{fig_chanmap}, which shows an integrated 
spectrum together with a set of binned channel maps. 
A complete set of channel maps is shown in Fig.\,\ref{fig_chanmap2}, 
wich is available in electronic form
at the CDS and can be obtained via anonymous ftp to cdsarc.u-strasbg.fr (130.79.128.5)
or via http://cdsweb.u-strasbg.fr/cgi-bin/qcat?J/A+A/.

The NW lobe of the outflow is pointing towards us (blue), the SE lobe is
pointing away.  This is
consistent with the Doppler shift of the Herbig-Haro object
HH\,494, located 6.15 arcmin northwest of CB\,26 on the axis of the
molecular outflow \citep[Fig.\ \ref{fig_overview};][]{stecklum_etal2004}.  
At the central position, emission from the disk and outflow are superimposed. 
The total projected
length of the bipolar molecular outflow is 14\arcsec, which
corresponds to $\approx$\,2000\,AU at 140\,pc.  However, the
separation between the CB\,26 disk and HH\,494 of $\approx 0.25$\,pc
suggests the total extent of the outflow is much larger.

\begin{figure*}
\begin{center}
\includegraphics[width=1.0\textwidth]{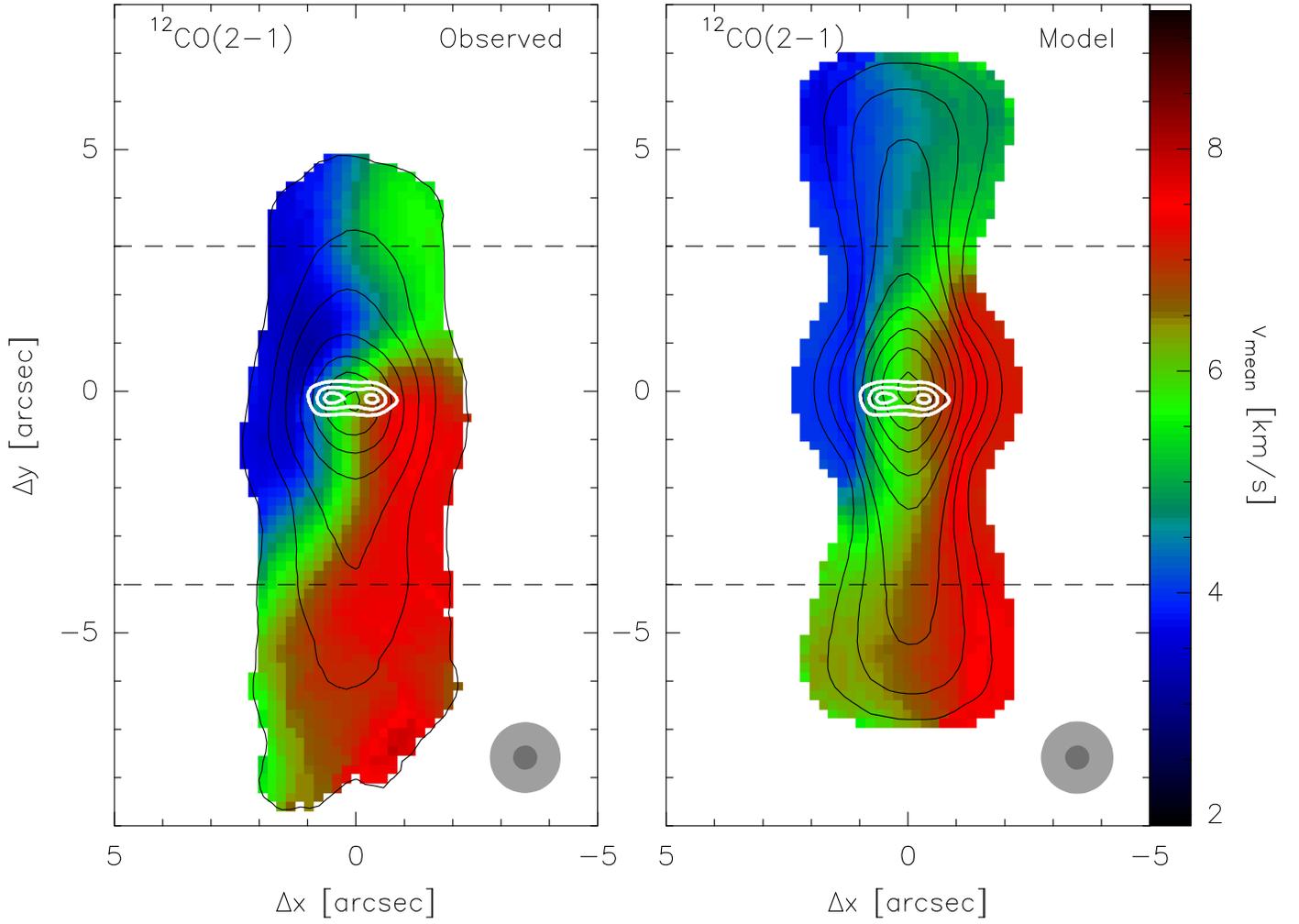}
\caption{\label{fig_12covelfield}
 $^{12}$CO\,(2--1) integrated intensity maps (contours) and mean velocity field 
(1$^{\rm st}$\ moment map, color) of CB\,26, rotated by 30\degr. 
 White contours show the 1.1\,mm dust continuum emission from the disk as observed with the SMA 
 (contour levels same as Fig.\,\ref{fig_overview}).
 The $^{12}$CO synthesized beam size is shown as large grey ellipse.
 The smaller and darker ellipse shows the 1.1\,mm continuum beam. 
 Left panel: observations.
 Right panel: best-fit model for $^{12}$CO\,(2--1). 
 Dashed lines refer to the y-coordinate of the position-velocity diagrams 
 shown in Fig.\ \ref{fig_12co_pvhor}.}
\end{center}
\end{figure*}

\begin{figure*}
\begin{center}
\includegraphics[width=1.0\textwidth]{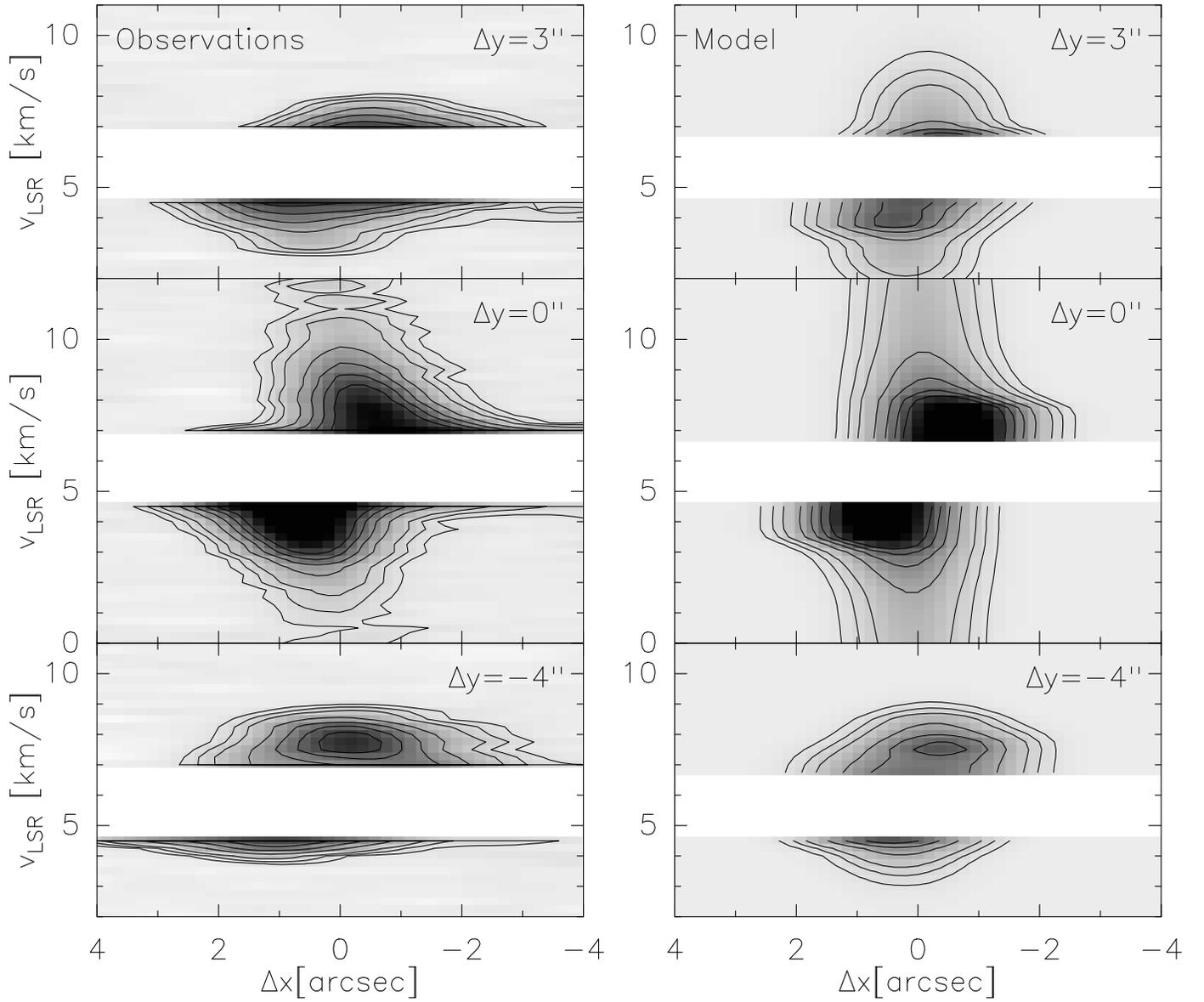}
\caption{\label{fig_12co_pvhor}
  Position-velocity diagrams of $^{12}$CO\,(2--1) perpendicular to the jet axis. 
  The coordinate along the jet axis 
  ($\Delta y$) is written in top right corner of each panel. 
  Contours start at 60\,mJy/beam (3\,$\sigma$). 
  Left panel: observations.
  Right panel: best-fit model}
\end{center}
\end{figure*}

Both the 1$^{\rm st}$\ moment map of the $^{12}$CO\,(2-1) emission
(Fig.\,\ref{fig_12covelfield}) and the position-velocity diagrams
(Fig.\,\ref{fig_12co_pvhor}) indicate the presence of a systematic
velocity gradient perpendicular to the flow axis along the entire
outflow.  The orientation of this lateral velocity gradient is the
same as that in the disk. 
The central position-velocity diagram
across the disk shows the classical Keplerian velocity pattern
\citep[Fig.\,\ref{fig_12co_pvhor}; cf.][]{lau_sar2001}.  The
position-velocity diagrams at +3\arcsec\ and $-$4\arcsec\ away from
the disk show clearly that the CO emission from the south-west side of
the outflow axis (negative $\Delta x$) is red-shifted with respect to
that from the north-east side (positive $\Delta x$) by about 2\,$\kms$.  
The amplitude of this lateral velocity gradient is comparable to 
that of the longitudinal velocity shift due to the bipolarity of the outflow. 
Although this is probably a pure coincidence due to the high inclination angle 
with respect to the line of sight (see \S\ \ref{ssec_mod-disk}), 
it has certainly helped to disclose these kinematic features in the first place 
(see also discussion in \S\ \ref{ssec_dis-rot}).
The observed radial velocity field of CB\,26 is very suggestive of a {\it
rotating outflow}.
To test this hypothesis and to derive more
quantitative results, we develop in the following section an empirical
model of the disk-outflow system and compare it to the full
$^{12}$CO(2--1) data cube.

\section{Modeling disk and outflow} \label{sec_mod}

In order to reproduce the observed $^{12}$CO\,(2-1) maps of CB\,26,
models of both the disk and the outflow are required. In this paper,
we use a fixed disk model that was derived from dust continuum
observations and develop an empirical model for the outflow. 
We then compute synthetic CO maps 
for the combined disk$+$outflow system and compare them 
to the observations (on the $T_{\mathrm mb}$\ scale).
Note, that in the following we use two different notations for the 
radius vector: 
``$R$'' refers to cylindrical coordinates, while 
``$r$'' refers to spherical coordinates 
(see Fig.\,\ref{fig_sketch}).

\subsection{Disk model} \label{ssec_mod-disk}

For the disk, we use the best-fit empirical model derived from
multi-wavelength continuum observations of CB\,26 (Sauter et al. in
prep.).  The basis of this model is a Keplerian disk with the
following density distribution:
  \begin{equation}
 \rho_{\rm disk} = \rho_0 \left(\frac{R_0}{R}\right)^{\alpha_\rho}
 {\rm exp}{\left[-\frac{1}{2}\left(\frac{z}{h(R)}\right)^2\right] }
\quad {\rm with} \quad
 h(R)=h_0 \left(\frac{R}{R_0}\right)^{\alpha_h}
  \end{equation}
Here, $R$\ and $z$\ are the cylindrical coordinates and $R_0
=100$\,AU.  Furthermore, the model includes an optically thin extended envelope
to account for the scattered light observed at NIR wavelengths.  The
disk is assumed to be passive and is heated by the central star
only. Values of the free parameters ($\rho_0$, $\alpha_\rho$, $h_0$, and
$\alpha_h$) were derived by radiative transfer simulations with the MC3D
code \citep{wolf_etal2003} to fit the observed SED and the spatially
resolved maps at JHK NIR-bands as well as at three millimeter
wavelengths.  
In particular, the following parameters were derived: 
$\alpha_\rho = 2.2$,
$h_0 = 10$\,AU, and 
$\alpha_h = 1.4$.
The disk mass remains somewhat uncertain, since the grain properties, 
and hence the dust opacities, are not well-constrained.
Assuming that this young disk (see
\S\,\ref{ssec-modres-star}) has a ``standard'' interstellar
gas-to-dust ratio, its total mass would be $\geq\,0.1\,\msun$. 
This mass is consistent with the simple estimate derived directly from the 1.3\,mm 
continuum flux of 190\,mJy and using 
$\kappa_{\rm 1.3mm} = 1.0$\,cm$^2$\,g$^{-1}$\ of dust \citep{oss1994}, 
a ``standard'' gas-to-dust ratio of 100,
and $\langle T_{\rm d}\rangle \approx 20$\,K. 
The exact value of the disk mass has very little impact on our result, 
as the CO emission is optically thick.

The disk was found to have an inner hole of radius
$\approx$\,45\,AU. Note, that the 1.1~mm continuum image presented in
Fig.~\ref{fig_overview} shows two peaks, which were already strongly
suggesting the presence of an inner gap. The disk modeling also
indicates an outer radius of 200\,AU.  

The adopted disk model was combined with line radiative transfer (LRT)
analysis of the HCO$^+$\,(1-0) and $^{12}$CO\,(2-1) Keplerian velocity
profiles to derive the dynamical mass of the central star(s) and the
inclination of the disk, $M_*=0.5\pm 0.1\,\msun$\ and 
$i = 85\pm 4^\circ$, respectively.
This mass is somewhat higher, but still comparable to the value 
of $\approx 0.3\,\msun$\ derived by \citet{lau_sar2001} from 
$^{13}$CO\,(1-0) only.

The $^{12}$CO\,(2--1) emission of this disk was then computed and
added to the outflow model described below, to be compared to the
observations.  We want to note that the parameters derived for the outflow do not
critically depend on the details of the adopted disk model. A much
simpler disk model, which we used first, lead to the same conclusions
for the outflow.

\subsection{Outflow model} \label{ssec_mod-of}

\begin{figure*}
\begin{center}
\includegraphics[width=1.0\textwidth]{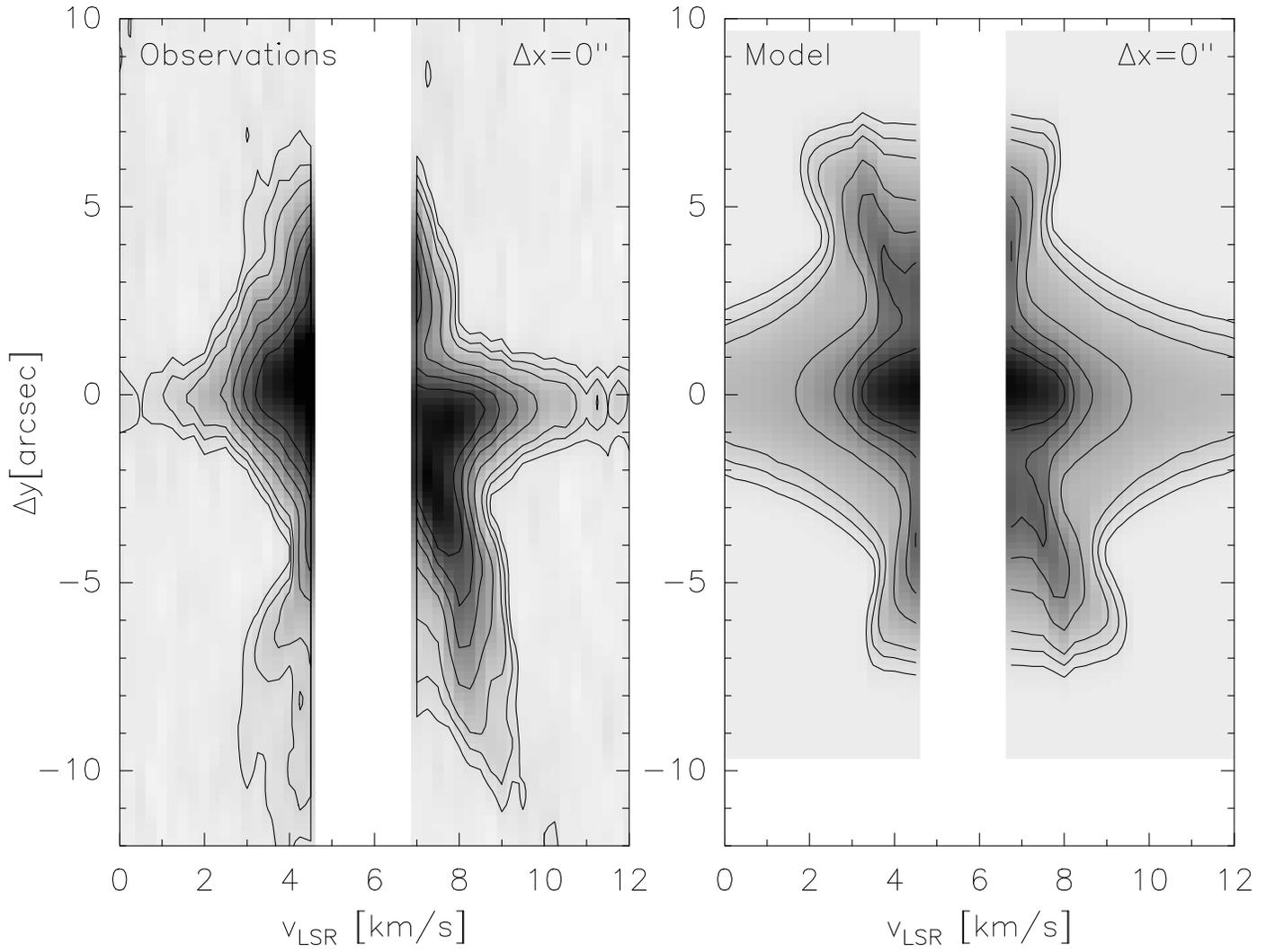}
\caption{\label{fig_12co_pvvert}
  Position-velocity diagrams of $^{12}$CO\,(2--1) parallel to the jet axis 
  at $\Delta x = 0$\arcsec. 
  Contours start at 60\,mJy/beam (3\,$\sigma$). 
  Left panel: observations.
  Right panel: best-fit model}
\end{center}
\end{figure*}

\begin{figure}
\begin{center}
\includegraphics[width=0.5\textwidth]{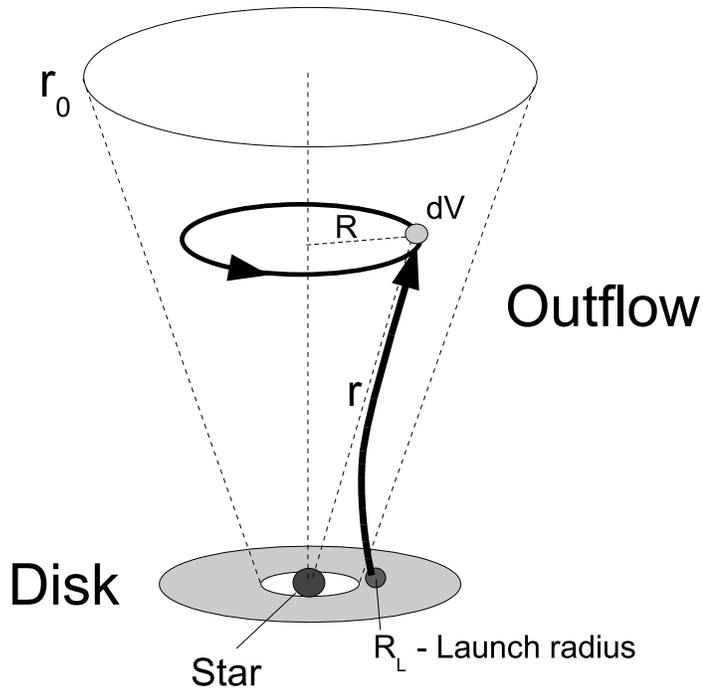}
\caption{\label{fig_sketch}
Schematic sketch of the outflow model. 
dV is the selected element
of the outflow, $r$\ and $R$\ are the radial and cylindrical
coordinates of the selected element, respectively. The element is launched
from the disk at radius $R_L$. $r_0$ is the length
of the outflow (see \S\,\ref{ssec_mod-of}). Matter flow lines (schematic) are 
shown as solid lines with arrows, while pure geometric connecting lines are dashed.}
\end{center}
\end{figure}

As there exists no unified physical paradigm for molecular outflows,
their prescription still relies on {\it empirical} parameterized models 
adapted to the particular source. To model the CB\,26 outflow, we use a
conical geometry of maximal size $r_0$\ (see Fig.\,\ref{fig_sketch}). To
reproduce the shape of the position-velocity diagram along the jet axis
(Fig.\,\ref{fig_12co_pvvert}), we assume that the radial outflow
velocity is a linear function of the distance from the star:
$V(r)=V_{0}\cdot r/r_0$. 
Such ``Hubble'' law-type velocity fields have been observed in many 
molecular outflows \citep[e.g.,][]{lada_fich_1996}, but their origin 
is not yet well-understood \citep[see also][]{DC_2003}. 
We further consider a steady-state model: at
any radial distance $r$\ from the central star, 
$r^2 n(r)V(r) = r_0^2 n_0 V_0$, where $V_0$\ and $n_0$\ are the radial
outflow velocity and density at distance $r_0$.  The radial H$_2$\
density distribution is then given by $n(r) = n_0\cdot(r/r_0)^{-3}$.
For the distribution of the kinetic gas temperature in the outflow we
assume a power law: {$T(r)=T_{0}(r/r_0)^{-q}$}.

We assume that the outflow is rotating around its polar axis.
In principle, one can consider different prescriptions for the 
radial dependence of the rotation velocity of the outflow. 
We chose a simple prescription, $V_{rot} \propto R^{\mu}$, and
modeled the following cases: 
$\mu = 1$\ (solid body rotation),
$\mu = -0.5$\ (Keplerian rotation), and 
$\mu = -1$\ (angular momentum conservation).
However, the data did not allow
us to reliably distinguish between these cases and
to further constrain the parameter $\mu$.
Therfore, we adopted the physically most plausible scenario in this range. 
We assume that the rotation velocity is given by $V_{rot}(R) = V_{kep}(R_L) \cdot R_L/R$,
where $R$\ is the distance from the outflow axis and $V_{kep}(R_L)$\ is the Keplerian
velocity at $R_L$ (see Fig.\,\ref{fig_sketch} and discussion in \S\,\ref{ssec_modres-unc}). 
This prescription is equivalent to the conservation
of the angular momentum along the flow, if all material was ejected from the disk
at distance $R_L$\ with the tangential component of ejection
velocity equal to $V_{kep}(R_L)$.

Finally, we added a microturbulent velocity of 0.1\,km/s, which is a
typical value for protoplanetary disks \citep{Guilloteau_Dutrey1998}
and assumed that CO is not depleted and has a constant abundance of
X(CO)\,=\,$7.5\cdot 10^{-5}$\ (see \S\,\ref{ssec_modres-unc}).
Synthetic $^{12}$CO\,(2-1) channel maps were produced using the LRT
code URAN(IA) \citep{pav_2007},
taking as input both the outflow and the disk models predictions.  The
free parameters $V_0$, $R_L$, $n_0$, $T_{0}$, and $q$ of the
outflow model were then derived by fitting the synthetic maps to the
observations. The computed maps were convolved with a Gaussian beam of
1.5\arcsec\ and compared with the observed channel maps using the
$\chi^2$ criterion:
\begin{equation}
\chi^2=\frac{1}{N\,\delta I^{obs}}\sum_{pos,V}\left(I^{mod}_{pos,V}-I^{obs}_{pos,V}\right)^2.
\end{equation}
Here, $I^{mod}_{pos,V}$ and $I^{obs}_{pos,V}$ are the modeled and
observed intensities for a given position (pixel) and velocity channel,
$\delta I^{obs}$ is the noise level of the observed intensity in the
channel maps, and summation is made over all positions and velocity
channels in the observed map ($N=N_{pos}\times N_{V}$). Because of the
large number of free parameters, we performed the
$\chi^2$-minimization in several 2D cuts of the parameter space. This
does not guarantee the exact localization of the minimum, but gives a
reasonable approximation to it. This approach is justified since our
goal was not to derive precise outflow parameters but rather to
illustrate that the observed CO emission can indeed be described by a
rotating outflow.

\section{Results} \label{sec-modres}

\subsection{The disk and central star(s)} \label{ssec-modres-star}

Although we have no direct indication for a double star or double jet
in CB\,26, the large inner hole in the disk of $R_{\rm in}\sim 45$\,AU
(see \S\,\ref{ssec_mod-disk}) suggests the presence of a binary
system. Assuming that this hole is due to tidal truncation by a binary system 
with approximately equal masses, the components should have a
separation of $a\sim 20$\,AU \citep[see][]{artym_1994}.
As noted above, the Keplerian profiles of the line emission
indicate a total stellar mass of $0.5\pm 0.1\,\msun$. 

In a recent paper, \citet{Guilloteau_etal2008} have derived very similar
properties (stellar mass, size of the central gap) for the HH\,30 disk. 
Combining with the results of the optical jet study of 
\citet{Anglada_etal2007}, they also infer the presence of a binary system with 
separation $a\sim 15$\,AU, orbital period $P\sim80$\,yrs, and 
total mass $\sim0.5\,\msun$, although the mass ratio remains poorly constrained.  

Integrating the spectral energy distribution (SED) of CB\,26 yields a
total luminosity of $L_{\rm bol} \ge 0.5$\,L$_{\odot}$\
\citep{stecklum_etal2004}. This can only be a lower limit to
the intrinsic luminosity since the edge-on view and bipolar geometry
cause an anisotropic radiation field where only a fraction of the NIR
photons are scattered into the line of sight.  Higher intrinsic
stellar luminosities of up to $\sim 2$\,L$_{\odot}$\ are consistent
with the model (but imply slightly different morphologies of the disk
and envelope). Our model corresponds to $L_{\rm bol} =
0.9$\,L$_{\odot}$.  Nevertheless, this luminosity range does allow an
estimate of the age of the system.  Using solar-metalicity PMS tracks
of \citet{Siess_etal2000} and a star mass of $0.5\,\msun$, we derive
an age of 1\,Myr, with lower and upper limits of 0.6\,Myr and 2\,Myr,
respectively. Assuming two stars with $0.25\,\msun$\ each, we derive a
similar age range, but the lower limit shifts to 0.1\,Myr.  However,
given the well-developed disk and the relatively thin remnant
envelope, such a young age does not seem realistic. We can therefore
safely constrain the age of CB\,26 to be $\approx$1\,Myr with an
uncertainty of factor 2.

\subsection{The outflow} \label{ssec-modres-of}

Table\,\ref{tab-mod} summarizes the parameters of the best-fit outflow
model. The radial velocity field of the model can
be compared to the observations (Fig.\,\ref{fig_12covelfield}): it
reproduces well the mean radial velocity pattern (red and blue areas) and
the S-like shape of the zero offset velocity emission (green area).
Furthermore, the computed position-velocity diagrams show the same 
radial velocity shifts across
the outflow as the observed data (Fig.\,\ref{fig_12co_pvhor}).
Although we can qualitatively reproduce all significant features which
are indicative of outflow rotation, the model fit is not perfect.  In
particular, we can not reproduce the asymmetry of the outflow along
its axis and the exact shape of the position-velocity diagrams.  We
believe that this is a drawback of our simplified and symmetric model.

The mean H$_2$ density in the outflow at $r_0 = 1000$\,AU is
$2\times$10$^4$~cm$^{-3}$.  The kinetic gas temperature at this radius
is $\ge$15\,K. The $^{12}$CO\,(2--1) emission is moderately optically thick 
throughout most of the outflow.
The derived radial outflow velocity at 1000~AU of 10\,$\kms$\
is comparable to that of HH\,30 \citep{Pety_etal2006}, but has a large
uncertainty (up to an order of magnitude) because of the uncertain
inclination (see below). The rotation velocity
is $\sim$1\,$\kms$ at a distance of 100~AU from the ejection axis.

The dynamical timescale of the molecular outflow 
(ratio of projected length and mean radial velocity, independent of the inclination)
is $\tau_{\rm dyn}\approx$ 500\,yr.
This is much shorter than the estimated age of the source of $\approx 1$\,Myr 
(\S\,\ref{ssec-modres-star}).
However, the presence of the Herbig-Haro object HH\,494 at a much larger distance
\citep[Fig.~\ref{fig_overview} and][]{stecklum_etal2004}
indicates that the total size, and hence the age, of the outflow 
is much larger than what we currently see in the $^{12}$CO\,(2-1) line.

The total mass of the visible molecular outflow
of $\approx\,10^{-3}\,\msun$\ corresponds to about 1\% of the disk
mass.  Although it is at the low-energy end, the CB\,26 outflow fits
well on the observationally established source luminosity -- outflow
momentum flux relation for protostars presented by
\citet{Richer_etal2000}.

\subsection{Modeling uncertainties} \label{ssec_modres-unc}

Interferometric channel maps of molecular lines from circumstellar
disks can be used to constrain the inclination angle and derive the
kinematic mass of the central star
\citep{Guilloteau_Dutrey1998,Simon_etal2000}.  However, there is a
general degeneracy between disk inclination and star mass.  While for
face-on configurations the inclination angle can be well-constrained
and the star mass remains uncertain, the situation is opposite for
edge-on configurations such as CB\,26.  Using geometrical arguments,
we could however constrain the inclination angle of the CB\,26 disk to
be $\le$89\degr\ (assuming the radial outflow velocity is
$<$100\,km\,s$^{-1}$) and $>$80\degr\ (from the aspect ratio of the
1.3\,mm dust continuum emission).  Therefore, we adopted a value of $i
= 85^{\circ}\pm4^{\circ}$.  In this inclination range, the central
star mass is well-constrained by the Keplerian rotation field and we
derive $M_{\ast} = 0.5\pm 0.1\,\msun$.  However, the projection factor
$1/{\mathrm{cos}}\,i$\ has a total uncertainty of one order of
magnitude, which leads to correspondingly large uncertainties in
estimates of outflow energetics and timescales.

Another degeneracy exists between hydrogen density and CO abundance.
This would lead to corresponding uncertainties in the estimates of
outflow mass and angular momentum.  We cannot directly derive the CO
abundance from our data, but we argue that the mean kinetic gas
temperature in the outflow of $\geq$\,15\,K is high enough to release
all CO molecules from the grain mantles into the gas phase
\citep{Bisschop_etal2006} and assume maximum possible CO abundance.

A more severe uncertainty comes from the fact that the small number of
usable velocity channels (due to the effect of the envelope; see
\S\,\ref{sec_obs} and Figs.\,\ref{fig_chanmap}, \ref{fig_12co_pvhor}, 
and \ref{fig_12co_pvvert}) and the moderate SNR
of the data did not allow us to well-constrain the radial dependence
of the outflow rotation velocity, $V_{rot}(R)$.  The assumption of
$V_{rot} \propto R^{-1}$\ was made because it is the physically most 
plausible profile within the range of profiles that are compatible 
with the data. However, we also want to stress that the other outflow 
parameters and conclusions we derive in this paper do not critically 
depend on the exact radial profile of the outflow rotation velocity.

\section{Discussion} \label{sec_dis}

\subsection{Alternative explanations for the velocity field of the outflow} \label{ssec_dis-vel}

The previous sections presented the model of a rotating outflow, that
was successfully fitted to the data. However, we also considered a
number of other configurations that could possibly lead to the
observed peculiar velocity field of the CB\,26 outflow:
\begin{enumerate}
\item molecular clouds which intersect with the line of sight and mimic a
   rotation-like velocity pattern,
\item two mis-aligned (non-rotating) outflows that are not spatially resolved,
   and
\item a precessing (non-rotating) jet\,/\,outflow.
\end{enumerate}

Given the relatively isolated location of the small globule CB\,26
(see Fig.\,\ref{fig_overview}), the clear jet-like bipolar morphology
of the CO emission, and the single-component line shapes (see
Fig.\,\ref{fig_chanmap}), scenario (1) would require a very special
spatial and kinematic configuration of the environment and seems
extremely unlikely. 
The alignment of HH\,494 (P.A.\,=\,$145$\degr) with the 
axis of the CO outflow 
(P.A.\,=\,$147\pm 3$\degr) also argues against this 
scenario (see Fig.\,\ref{fig_overview}).

We tried to model two (spatially unresolved) outflows that are
slightly misaligned with respect to the line of sight (scenario 2).
The suspected binarity of CB\,26 could make this scenario appealing.
However, until now, only very few close binary systems 
are known to drive two jets; e.g., 
there is evidence that the quadrupolar outflow 
associated with HH\,111 is driven by a close  
binary \citep[separation $<40$\,AU in projection;][]{rodriguez_2008}.
Two outflows from the same central core 
(the so-called quadrupolar outflows)
are usually observed only from sources with much larger separations
\citep[e.g., a few thousand AU in HH\,288 or CG\,30;][]{Gueth_etal2001,Chen_etal2008b}. 
Still, we
have modeled this configuration and found that it can also produce an
effective velocity gradient across the map of the two combined flows.
However, reproducing the S-shape of the zero-velocity zone in the
1$^{st}$\ moment map (green area in Fig.\ \ref{fig_12covelfield})
requires fine-tuning of both the relative orientation and the strength
the two flows. The most significant differences to a rotating
configuration occur near the driving sources and at low radial 
velocities.  Unfortunately, both the dominating disk emission at the
center and the missing low-velocity channels due to masking by the
non-recoverable envelope emission prevent us from ultimately
confirming or disproving this scenario as possible origin of the
peculiar velocity field of CB\,26.  Note, however, that the agreement
between the velocity gradient directions of disk and outflow would be
a pure coincidence in this scenario.

Entrainment by an unseen precessing jet (scenario 3) could possibly
also mimic a similar radial velocity field when the outflow traces about
0.25 to 0.5 of the precession period. 
In this case, we would not observe rotation, but see gas that is accelerated 
along different directions at different times.  However, longitudinal 
gradients would dominate the radial velocity field over lateral ones. 
Since simulating this scenario would require a non-stationary model, 
we have only considered it as a thought experiment. 
Assuming that the precession of the jet axis is driven by
tidal interactions between the circumstellar disk from which the jet
is launched and a companion star on a non-coplanar orbit, the shortest
possible precession period in CB\,26 would be of order 2000\,yr
\citep[][see their eq. 14]{Terquem_1998,Anglada_etal2007}.
Here we assumed equal companion masses, small inclination angle,
maximum ratio between circumstellar disk radius and companion
separation of 0.3, and an orbital period of $\approx$100\,yr
(\S\,\ref{ssec-modres-star}).  Comparing this to the dynamical
timescale of the molecular outflow of 500\,yr
(\S\,\ref{ssec-modres-of}) leads to the conclusion that at least the timescales 
are such that this scenario cannot be excluded to be at work for CB\,26.
It would, however, be worthwhile to test the precession scenario with an 
appropriate model.

\subsection{A rotating outflow -- what distinguishes CB\,26 from other sources?} \label{ssec_dis-rot}

Assuming the velocity pattern of the CB\,26 outflow does result
from a rotation of the outflowing material about its ejection axis,
the data presented in this paper would be the first observational
signature of this phenomenon.

If the angular momentum transfer from the disk by an (unseen) rotating jet drives
the rotation of the molecular outflow, the question that arises is 
why such a signature is so prominent in CB\,26 while it has
not been observed in other sources? Detecting the projected velocity
gradients requires an nearly edge-on configuration, which is the case
of CB\,26 but also of other sources that have been studied. Some
specificities of CB\,26 must therefore be invoked, but these are
difficult to spot without being too speculative.

The comparison between the CB\,26 and HH\,30 
\citep{Pety_etal2006,Guilloteau_etal2008} cases illustrates that difficulty. By
many aspects, these two sources are very similar: both systems are
seen almost edge-on, with inclination larger than $\sim$84\degr; the
disks are in Keplerian rotation around a $\sim$0.5$\msun$ system and
have an $\sim$40~AU inner gap; and both sources are driving compact
molecular outflows. 
Binarity seems the most simple explanation for both the mm and 
optical appearance of HH\,30 \citep{Anglada_etal2007,Guilloteau_etal2008}, 
while the binarity status of CB\,26 is still speculative. 
If CB\,26 is indeed a binary system, it is likely to
have similar characteristics to those of HH\,30, as the masses of the
central star(s) and the sizes of the cavities are similar. As for the 
CO outflow, the HH\,30 flow has a somewhat larger opening angle 
(60\degr) than CB\,26 (40\degr, see Table \ref{tab-mod}) 
and presents only one lobe (which remains a puzzling
feature, probably related to the environment of the protostar rather
than to an intrinsic property of the ejection mechanism). More
importantly, and despite a detailed analysis of the CO emission, the
HH\,30 outflow does not show any sign of rotation
\citep{Pety_etal2006}. If the velocity pattern observed in CB\,26 is
indeed rotation, this raises the question of why rotation is not
observed in a quite similar object like HH\,30? The most significant
difference between the two sources is their age: CB\,26 is likely to
be younger, since it has more prominent envelope.
However, there are no clear prescriptions or predictions on the evolution 
of jet/outflow rotation with time.

More generally, we note that the angular momentum transfer from the
central sources to the outflowing gas may be quite complex in a binary
system -- as suspected in CB\,26. One could, e.g., speculate on the
presence of two distinct parallel jets, that are creating one single outflow,
but ``stir'' the entrained gas and disperse {\em orbital} angular
momentum from the system
\citep[cf.][]{Masciadri_Raga_2002}.

\subsection{The relation between disk and outflow} \label{ssec_dis-diskof}

If we make the two simplifying assumptions that the outflow mass flux
is constant over time and that the outflow is fed directly by the
disk, it would take $\approx$0.5\,Myr to dissipate the disk (see
Table\,\ref{tab-der}). This compares well to the age of the system
and, within the uncertainty range, to the typical dispersal time of
circumstellar disks around young stars
\citep{Haisch_etal2001,Cieza_etal2007}.  We also note that this mass
outflow rate is comparable to typical accretion rates for young
T~Tauri stars \citep{Hartigan_etal1995,Gullbring_etal1998}.  Assuming
that the outflow indeed rotates, the average specific angular momentum
of the outflow gas is somewhat lower or comparable to that of the disk
material.  If we make the same assumptions for the angular momentum
dispersal as above, we derive a timescale of $\approx 1$\,Myr.  The
good agreement between both balance and timescale estimates supports
the hypothesis that the CB\,26 outflow is indeed rotating and
dispersing angular momentum from the system.  In any case, it also
shows that in CB\,26 the outflow plays a significant role in
dispersing the disk, and thus terminating the conditions for planet
formation.  

The rotation of the outflowing gas can also be used to put constraints on the
size of the ejecting region in the disk, assuming  we are observing gas
that has been directly ejected from the disk, and not envelope material
that has been accelerated by an underlying (unseen) jet. 
Using the simple assumptions that the outflow material is launched 
from a single point on the Keplerian disk and that angular
momentum is conserved, i.e., 
$R\cdot V_{rot}(R) = R_L\cdot V_{\rm kep}(R_L)$,
we derive from our $\chi^2$--fitting routine $R_{\rm L} \approx 30$\,AU. 
This is consistent with the simple estimation
\begin{equation}\label{eq-rl}
R_L=\frac{(V_{\rm rot} R)^2}{G M_{\ast}}~ ,
\end{equation}
which yields $R_{\rm L} \approx 27$\,AU when we assume 
$M_{\ast}=0.5\msun$, $R=100$~AU, and $V_{\rm rot}=1.1\kms$\ 
(see Table\ \ref{tab-mod}).

One can also use magneto-centrifugal wind models 
\citep[see review by][]{Shu_etal2000} to constrain the outflow launch 
region. In this theory, the magnetic field plays an essential role in extracting
the energy and angular momentum from the disk and redirecting them
to the outflow. \citet{Anderson_etal2003} have shown that there is 
a useful invariant that is conserved along a magnetic field line and
that can be used to determine the radius $R_{\rm ML}$ at which the
material was ejected. 
\citet{Pety_etal2006} have shown that this relation can be written as
\begin{equation}\label{eq-rml}
 R_{\rm ML} = R\left( 2\frac{V_{\rm kep} V_{\rm rot}}{V^2}\right)^{2/3}~ ,
\end{equation}
where $R$\ is the distance from the outflow axis to the considered volume element, 
$V_{\rm kep} = \sqrt{GM_*/R}$\ is the Keplerian velocity
at this position, and $V_{\rm rot}$ and $V$\ are the measured rotation and
radial outflow velocities at that position. 
However, the Hubble law-type radial outflow velocity field (see \S\,\ref{ssec_mod-of}) 
implies that the launching radius derived with this equation depends on the 
distance of the catching point from the star. 
For eq.\,\ref{eq-rml}\ to be applicable, the catching point should far away from 
the star \citep{Anderson_etal2003}. 
On the other hand, the Hubble law describes the observed radial outflow velocity 
field of CB\,26 reasonably well at small and intermediate distances from the star, 
while the outflow velocity appears to be more constant at larger distances 
(see Fig.\,\ref{fig_12co_pvvert}).
If we therefore use $V=5\kms$\ at $r=500$\,AU (see Table\ \ref{tab-mod} and 
\S\,\ref{ssec_mod-of}), and further assume $M_*=0.5\msun$, $R=100$~AU, and 
$V_{\rm rot}=1.1\kms$, we obtain $R_{\rm ML} \sim 32$\,AU. 
This value is remarkably close to the values obtained from the simple 
estimates above. Using $V=10\kms$\ at $r=1000$\,AU leads to $R_{\rm ML} \sim 13$\,AU, 
which can be considered a lower limit to $R_{\rm ML}$.
However, as all these estimates are based on a number of unproven assumptions, 
we do not use them to draw conclusions on the physical launch mechanism.

\section{Summary and Conclusions} \label{sec_sum}

We presented new millimeter-interferometric observations of the young
edge-on T\,Tauri\,star -- disk system in the isolated Bok globule
CB\,26.  Our $^{12}$CO(2--1) line data at 1\asp5 spatial and
0.25\,$\kms$\ velocity resolution reveal a previously undiscovered
collimated bipolar molecular outflow of total length $\approx
2000$\,AU, escaping perpendicular to the plane of the well-resolved
circumstellar disk in Keplerian rotation.  The disk has an inner hole
of radius $\approx$45\,AU, indicating the possible presence of a
binary star system with separation $\approx$20\,AU.  The dynamic mass
and age of the central obscured star(s) is $0.5\pm 0.1\,\msun$\ and
$\approx$\,1\,Myr, respectively.  

The outflow shows very clear and consistent kinematic signatures that
suggest it is rotating about its polar axis.  We use an empirical
parameterized outflow model combined with 2-D line radiative transfer
calculations and $\chi^2$-minimization in parameter space to derive a
best-fit model and compute parameters of the outflow.  We could show
that the data are consistent with outflow rotation.  This
hypothesis is supported by the fact that disk and outflow seem
co-rotating and fit together energetically.  However, we could not
ultimately exclude jet precession or two misaligned flows as possible
origin of the observed peculiar velocity field.

Assuming that the CB\,26 outflow is rotating, we computed and compared
mass, mass flux, angular momenta, and angular momentum flux of disk
and outflow and derived disk dispersal timescales of $0.5\ldots 1$\,Myr.
Although our simplifying assumptions and the model uncertainties do
not allow us to draw more precise conclusions on the outflow formation
and disk dispersal mechanisms, this result confirms the typical disk
dispersal timescale derived on statistical arguments.
It also shows that the CB\,26 outflow must
play a significant role in dispersing the disk.
 
Ultimately testing the proposed scenarios that could explain the 
peculiar radial velocity field of the CB\,26 molecular outflow 
will most likely require to combine the radiative transfer calculations 
with hydrodynamic models of jet formation and entrainment in multiple sources.



\begin{acknowledgements}
We acknowledge the Plateau de Bure and Grenoble IRAM staff for their help
during the observations and data reduction.
This work was supported by the ``Action sur Projets Physique
Chimie du Milieu Interstellaire'' (PCMI) from INSU / CNRS.
J.Sauter and S.Wolf provided us with their preliminary results
on the disk modeling and we benefitted from many fruitful discussions with them.
We also want to thank K. Schuster and C. Fendt for helpful discussions. 
We are grateful to G. Anglada for critically reading the manuscript and 
providing us with constructive comments that helped to improve the 
correctness and clarity of this paper.
\end{acknowledgements}



\bibliography{cid_cb26-bib}
\bibliographystyle{aa}



\begin{table*}{}
\caption[]{\label{tab-mod} Parameters of the outflow model for CB\,26.}
\begin{tabular}[t]{llll}  
\hline \hline \noalign{\smallskip}
Parameter                       & Symbol         & fixed & fitted \\
\noalign{\smallskip} \hline \noalign{\smallskip}
Outflow Size                    & $r_0$          & 1000~AU     & -- \\
Opening angle                   & $\alpha$       & 40~deg      & -- \\
Positional angle                & PA             & 30~deg      & -- \\
Inclination                     & $i$            & 85~deg      & -- \\
Microturbulent velocity         & $V_{\rm turb}$ & 0.1\,km\,s$^{-1}$   & -- \\
CO/H$_2$ abundance:             & X(CO)          & $7.5\times 10^{-5}$ & -- \\
H$_2$ density at $r_0$          & $n_0$          &  --         & $2\cdot 10^{4}$\,cm$^{-3}$ \\
Radial outflow velocity at $r_0$        & $V_0$          &  --         & 10~km/s \\
Temperature at $r_0$            & $T_0$          &  --         & 15\,K   \\
Temperature index               & $q$            &  --         & $0.5$  \\
%
Rotation velocity at $R=$100 AU & $V_{rot}(100)$ &  --         & 1.1~km/s \\
\noalign{\smallskip} \hline\noalign{\smallskip} 
\end{tabular} 
\end{table*}

\begin{table*}{}
\caption[]{\label{tab-der} Derived parameters of outflow and disk.}
\begin{tabular}[t]{lll}  
\hline \hline \noalign{\smallskip}
Parameter                       & Symbol         & value \\
\noalign{\smallskip} \hline \noalign{\smallskip}
Total outflow mass              & $M_{\rm CO}$       &  $1\times 10^{-3}\msun$ \\
Mass flux                       & $\dot{M}_{\rm CO}$ &  $2\times 10^{-7}\,\msun\,{\rm yr}^{-1}$ \\
Total outflow momentum          & $P_{\rm CO}$       &  $1\times 10^{-2}\,\msun\,{\rm km\,s}^{-1}$ \\
Momentum flux                   & $\dot{P}_{\rm CO}$ &  $2\times 10^{-6}\,\msun\,{\rm km\,s}^{-1}\,{\rm yr}^{-1}$\\
Total angular momentum          & $J_{\rm CO}$       &  $1.5\times 10^7\,\msun\,{\rm km^2\,s}^{-1}$ \\
Angular momentum flux           & $\dot{J}_{\rm CO}$ &  $3\times 10^{3}\,\msun\,{\rm km^2\,s^{-1}\,yr^{-1}}$ \\
Specific angular momentum       & $j_{\rm CO}$       &  $1.5\times 10^{10}\,{\rm km^2\,s^{-1}}$\\[1mm]
Disk mass~$^{\ast}$                  & $M_{\rm disk}$     &  $1\times 10^{-1}\msun$ \\
Disk angular momentum      & $J_{\rm disk}$     &  $3\times 10^{9}\,\msun\,{\rm km^2\,s}^{-1}$ \\
Disk specific angular momentum  & $j_{\rm disk}$     &  $3\times 10^{10}\,{\rm km^2\,s^{-1}}$\\
\noalign{\smallskip} \hline\noalign{\smallskip} 
\end{tabular} 
\newline
$^{\ast}$\ The disk mass is derived from the disk model of Sautter et al. (in prep.).
\end{table*}



\clearpage

\clearpage


\end{document}